# Spatio-Temporal Optical Coherence Tomography provides advanced imaging of the human retina and choroid

Egidijus Auksorius,[1,2,3] Dawid Borycki,[1,3] Piotr Wegrzyn,[1,3,7] Bartosz L. Sikorski,[4,6] Ieva Zickiene,[2] Kamil Lizewski,[1,3] Mounika Rapolu,[1,3] Karolis Adomavicius,[2] Slawomir Tomczewski[1,3] and Maciej Wojtkowski[*1,3,5]

[1]*International Center for Translational Eye Research (ICTER), ul. Skierniewicka 10a 01-230 Warsaw, Poland*
[2]*Center for Physical Sciences and Technology (FTMC), Saulėtekio al. 3, LT-10257 Vilnius, Lithuania*
[3]*Institute of Physical Chemistry, Polish Academy of Sciences, Kasprzaka 44/52, 01-224 Warsaw, Poland*
[4]*Department of Ophthalmology, Nicolaus Copernicus University, Skłodowskiej-Curie 9, 85-090, Bydgoszcz, Poland*
[5]*Faculty of Physics, Astronomy and Informatics, Nicolaus Copernicus University, Grudziądzka 5, 87-100 Torun, Poland*
[6]*Oculomedica Eye Research & Development Center, Ogrody 14, 85-870, Bydgoszcz, Poland*
[7]*Faculty of Physics, University of Warsaw, Pasteura 5, 02-093 Warsaw, Poland*

*Corresponding author e-mail address: mwojtkowski@ichf.edu.pl*

**Abstract:**
One of the greatest needs in modern ophthalmology is access to the eye's choroid in vivo. Without it, it is difficult to introduce new therapies and to understand most of the pathological changes in the eye, because 85% of the total blood flow in the eye is caused by the choroidal circulation. It is entirely responsible for the metabolism of photoreceptors, which each of us feels every day and senses when its deficits begin. Optical selection of choroid and choriocapillaris by currently used imaging techniques such as indocyanine angiography, scanning laser ophthalmoscopy or optical coherence tomography (OCT) is limited by insufficient lateral resolution, depth of penetration and the presence of strong light scattering effects. Here we report a Spatio-Temporal Optical Coherence Tomography (STOC-T) that maintains microscopic quality of *in vivo* reconstructions of anatomical layers when imaging chorioretinal cross-sections in a coronal anatomic plane. STOC-T performs high definition volumetric imaging in all projections, enabling reconstruction of anatomical details of the retinal and choroidal layers with unprecedented quality by using an extra phase randomization for low spatial coherence. We also demonstrate that STOC-T can detect blood flow and reveal vascular networks in various chorioretinal layers that are otherwise invisible to OCT. This capability highlights the unique value of STOC-T angiography. These technical advances enable acquisition of detailed choroidal structures, such as Satller's layer and choriocapillaris, that are usually buried behind speckle noise and blurred by eye motion artifacts in conventional OCT. By imaging the retina and choroid with high contrast, we were able to perform detailed morphometric analysis, which is the basis for the anticipated clinical utility of the method in monitoring disease progression and therapeutic responses of the eye.

Age-related macular degeneration (AMD), diabetic retinopathy (DR), and glaucoma, among others, are debilitating diseases of the eye involving retinal dysfunction that leads to loss of visual acuity and eventual blindness, representing an extensive medical and social burden. Currently, early detection, halted progression, or reversal of the retinal pathology are not clinically possible, but advancing technology is closing the gap.

Optical imaging technologies have significantly improved visualization of retinal pathophysiology and have had a substantial impact on basic and translational medical research. These technologies, such as scanning laser ophthalmoscopy, and fluorescein and Indocyanine Green (ICG) angiographies have translated into more accurate diagnosis, and improved management of numerous diseases of the eye. In particular, Optical Coherence Tomography (OCT)[1,2] has become a standard of care for diagnosing and monitoring the treatment of such retinal diseases as AMD, DR, and glaucoma[3-7]. Thanks to its non-invasiveness and fast image acquisition, OCT has dramatically intensified research activities by enabling studies on a larger group of patients and in a shorter period of time than before. The advent of OCT Angiography (angio-OCT) provided a further boost in the field by allowing visualization of the retinal and choroidal microvasculature without the injection of exogenous dyes[8-10]. Indeed, the clinical use of OCT has become so widespread that the mass of images available have exceeded the ability to analyze them, so that artificial intelligence (AI) approaches have been developed to address this need[11]. However, AI is only so valuable as the quality of the input images, and there are shortcomings in the available OCT data. Thus, currently available OCT reconstructions of the anatomical layers of the retina in the most intuitive *en-face* projections do not provide information in sufficient detail to be of substantial clinical utility. This is especially true for both the retinal and the choroidal layers underlying the retinal pigment epithelium (RPE). It is still challenging for classical OCT techniques to distinguish morphological elements such as internal limiting membrane (ILM) structure, individual nerve fibers, nerve cell soma, photoreceptor and RPE mosaic, or choriocapillaris (CC) and other choroidal layers. The quintessential challenge of OCT imaging is that of the CC because of its fine structure, and interference from the pigmentation of the RPE layer and choroidal melanocytes that strongly absorb and scatter light. Abnormalities in CC circulation are critical in AMD[12], DR[13], glaucoma[14] and others. Therefore, detailed visualization of this complex is very important in both the research and clinical settings.

Angio-OCT has been used extensively to image the CC [15] but, unlike the retinal microvasculature, the CC is a much denser capillary network and provides less signal. Moreover, *en face* projections necessary for angio-OCT and derived from the data acquired with conventional-scanning OCT, normally lack the spatial resolution and contrast to adequately image the CC due to motion artifacts, speckle noise and ocular aberrations. The involuntary side-by-side eye movements during raster scanning of a laser point across the choroid heavily distorts images in conventional OCT. Even the fastest scanning OCT systems[16] cannot "freeze" the retina to eliminate motion artifacts. Furthermore, spatially coherent lasers, necessary for implementing the confocal detection, introduce a granular appearance in the OCT images, known as speckles, which arise from the interference of light backscattered from different retinal structures[17]. Speckles prevent the observation of microscopic structures in the chorioretinal images, even for those structures physically larger than the average speckle (*e.g.*, small capillaries), and this problem becomes worse for deeper layers, such as the outer retinal and choroidal layers. To visualize the retina *in vivo* without motion artifacts researchers have continued to push the speed limits of OCT[16,18]. In particular, systems with wide-field retinal illumination and backscattered signal detection with ultrahigh-speed cameras[19,20] are capable of reaching very high imaging speeds[21], and do not feature beam scanning artifacts met in conventional OCT. However, a serious drawback of such systems has been crosstalk noise that has prevented imaging the choroid.

In an attempt to visualize choroidal structures, and thus, improve the imaging depth of fast wide-field systems, we recently utilized phase randomization[22] to eliminate crosstalk noise in the retina[23], resulting in a class of imaging techniques that we term Spatio-Temporal Optical Coherence Tomography (STOC-T). We have also demonstrated that the speckle size in retinal images recorded with STOC-T can be three times smaller compared to conventional OCT[23], potentially making small details in the retina and choroid more discernible than in conventional OCT. Indeed, the technical advances introduced in this contribution enable imaging of human retina and choroid *in vivo* with high contrast. In particular, choriocapillaris is clearly reconstructed with good delineation of microscopic capillary system that lends itself to rigorous quantitative morphometry.

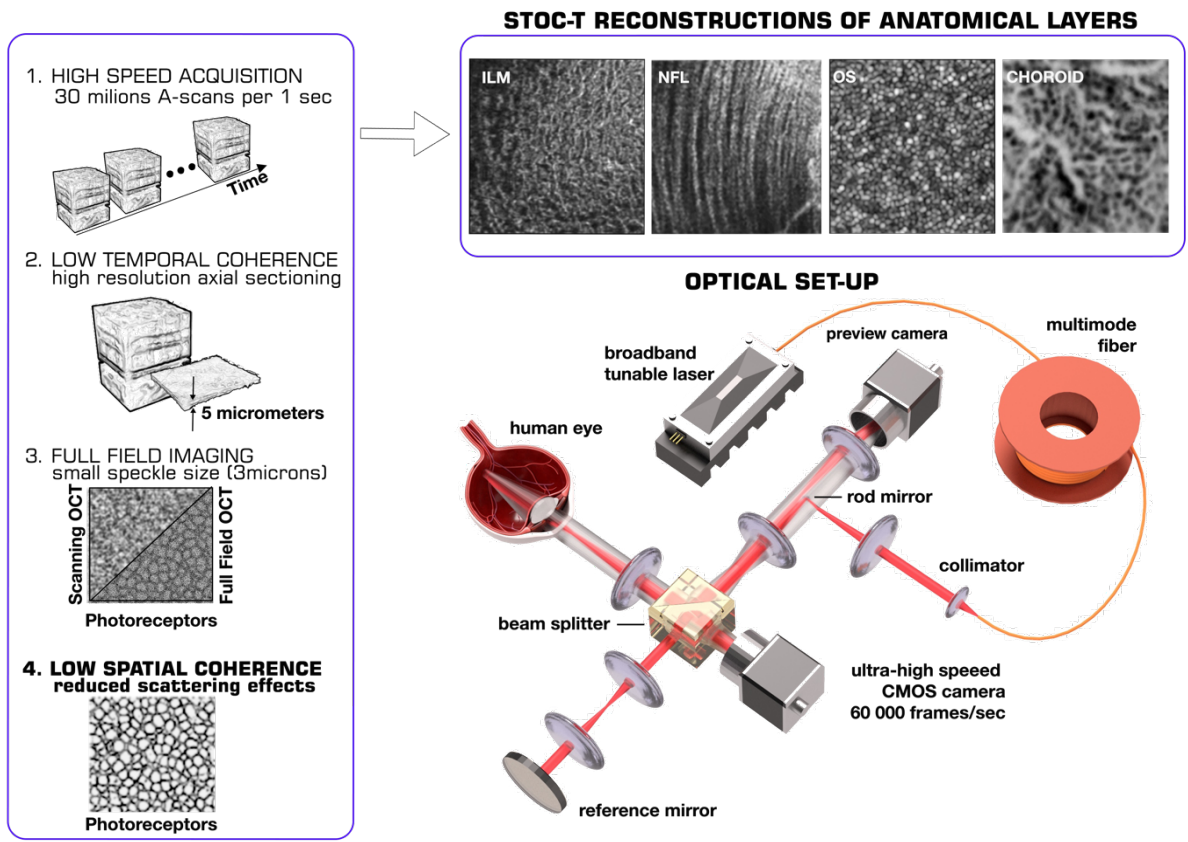

**Figure 1.** Spatio-Temporal Optical Coherence Tomography (STOC-T) principle (left) and scheme (right). **Left:** In STOC-T the retina is probed with complex spatiotemporal light fields that help produce retinal volumes in milliseconds with high-quality and highly-contrasted retinal layers. Signal detection from the retina with a camera allows high-speed imaging with small speckle size and reduced motion artifacts. Specifically, in conjunction with a tunable laser, it is capable of collecting 80 volumes ($512^3$ voxels corresponding to 1.7 mm x 1.7 mm x 1.7 mm) in a second, equivalent to a 30 MHz A-scans rate. Moreover, multimode optical fiber and wide-field detection enable retinal imaging with reduced detrimental scattering effects such as crosstalk and speckle noise by eliminating the former and reducing the speckle size by a factor of three. Finally, point-scanning artifacts are absent in STOC-T imaging, while motion artifacts can be successfully managed in postprocessing. **Right:** STOC-T is based on a simple interferometer design that incorporates an ultrahigh-speed camera and a tunable laser source for recording retinal volumes (in the form of multispectral interferograms). A 300 meter-long multimode fiber

with a 50 μm core produces ~400 well-defined spatial modes within nanoseconds that helps to quickly and efficiently remove crosstalk noise in wide-field images. A separate inexpensive camera provides video-rate axial images of the retina that is utilized to find focus on the retina with STOC-T. It detects backscattered signal from the retina that is lost in a conventional interferometer design (see Online Methods).

Figure 1 summarizes the unique combination of features in STOC-T that allows detailed and high-contrast imaging of the chorioretinal layers and shows a constructed STOC-T system with an efficient and fast phase randomization that is based on the use of a multimode fiber. In addition, the system was designed to rapidly acquire axial retinal images that were used as a visual feedback for a quick eye alignment and selection of imaging location. Such visualization greatly facilitated capturing volumes with the main (2-D) camera over multiple retinal locations that could be then stitched together to offer a large field of view of the retina.

## Results

The chorioretinal images were acquired *in vivo* with the STOC-T system, from a healthy 45-year-old volunteer. Multiple volumes (10 - 30) were acquired per single imaging location of 1.7 mm x 1.7 mm in size within 260 msec. Montage images of 9 mm x 4.6 mm in size were produced by stitching together volumes captured from 32 adjacent locations. We have also acquired complimentary time-resolved ICG angiography (Fig. S 13), Spectralis OCT (Heidelberg Engineering) and Triton OCT (Topcon) (Fig. S 11 and Fig. S 14) images from the same eye for comparison purposes.

### Retinal microstructure

Figure 2 demonstrates STOC-T capabilities to image the retina with high quality, and extract information about anatomical layers. It shows *en face* and axial images over an extended field of view produced by stitching volumes/images in the lateral plane. STOC-T offers visualization of anatomical details of the retina in *en-face* projection without the need for complex hardware solutions, resulting in effects previously unattainable for OCT. Especially spectacular is the imaging of the structure of the thin Inner Limiting Membrane (ILM) that separates the nerve fiber layer (NFL) from the vitreous body. Figure 2c and f show the structure of this membrane, mapping the architecture of individual nerve fibers located just below it (Figure 2d). In locations proximal to the optic disc, the ILM is more difficult to distinguish and the nerve fibers themselves begin to dominate. Figure 2f shows where the distinguishable ILM layer and nerve fibers are transitioning. Structural reconstructions of anatomical retinal layers reveal superficial, deep, and intermediate capillary plexuses (ICP and DCP in Figs. 2f and 2g and SDC fig. S16) without the need for OCT angiography. Additionally, the granular structure of the RPE and IS/OS can be clearly seen. With the additional use of a software-based technique of numerical aberration correction it is possible to optimize the transverse resolution to diffraction-limited and visualize the photoreceptor mosaic (Figure 2g insets)[24].

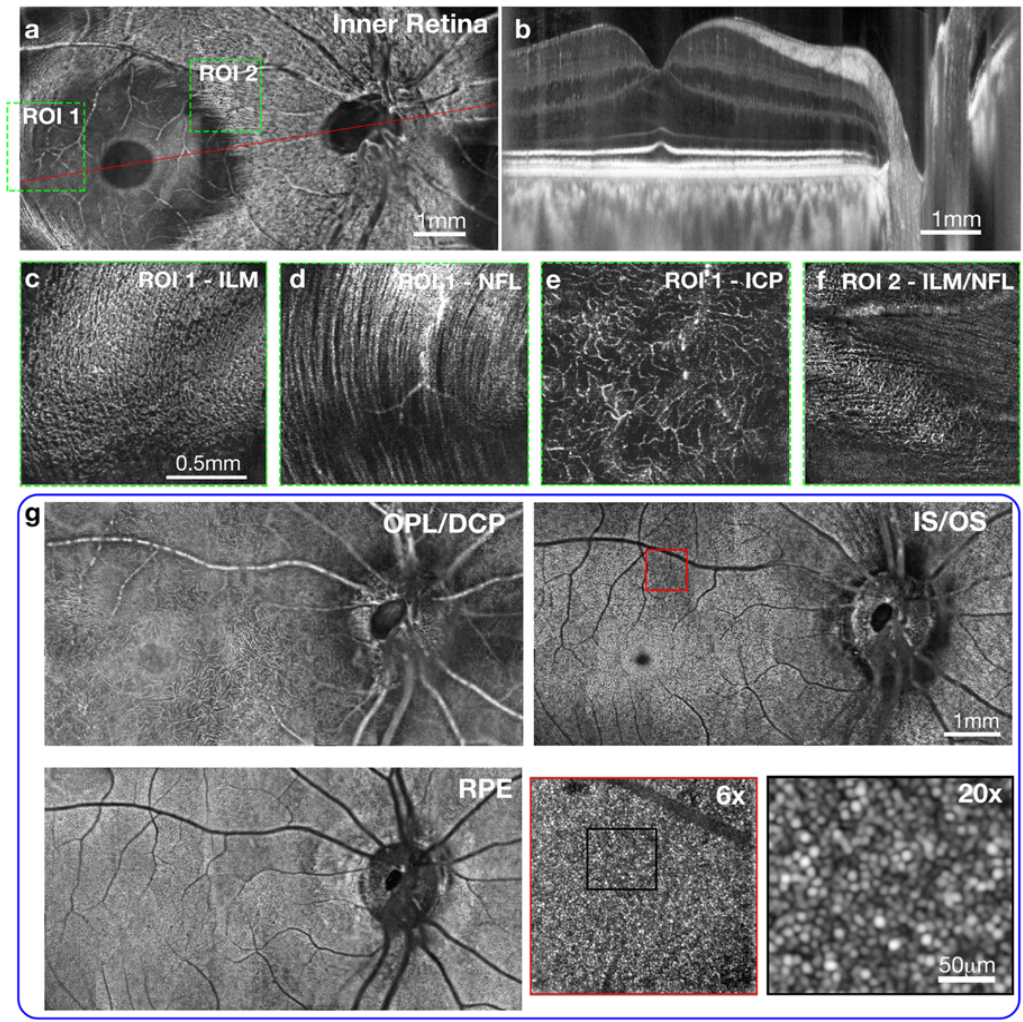

**Figure 2.** STOC-T imaging of retinal microstructure: montaged (stitched) retinal images (9 mm x 4.6 mm) of macula and optic nerve head; and panels with zoomed-in regions. **a.** 15 μm *en face* projection of the inner retinal layers featuring the optic nerve discs, retinal nerve fiber layer (NFL), and retinal ganglion cell layer (GCL) in one image. **b**. Cross-sectional image along the papillomacular axis of the human retina indicated by red line in panel a.; **c.**, **d., e.** Details of the inner limiting membrane (ILM), NFL and intermediate capillary plexus (ICP) are shown as thin layers (of 5 μm) and zoomed-in for the locations indicated by the dashed green boxes in a. **g.** STOC-T reconstructions of anatomical retinal layers (15 μm thick coronal projections): outer plexiform layer and deep capillary plexus (OPL)/DCP, inner segment/outer segment (IS/OS) junction and retinal pigment epithelium (RPE). Red box in IS/OS image is shown zoomed-in to various degrees below. The IS/OS reconstruction was numerically compensated for ocular aberrations revealing photoreceptor structure (22$^{nd}$ order Zernike polynomials).

## Choroidal microstructure

The choroid plays a key role in the proper function of the retina by nourishing its outer layers and transporting metabolic waste from the RPE[25]. The choroid is traditionally divided into three parts located from the outer to the inner part: Haller's layer, Sattler's layer, and the choriocapillaris, shown in Figure 4b. Sattler's layer contains the medium-sized vessels, and Haller's layer contains the larger vessels of the choroid. However, there is no clear boundary between these layers and, in addition, the choroidal layers experience significant individual variability[26]. The thickness of these layers changes from case to case, and also it depends on the location relative to the central fovea; typically it is thicker in the subfoveal zone whereas it thins considerably in the optic disc region. For this reason, there is no universal value for the thickness of Sattler's and Haller's layers.

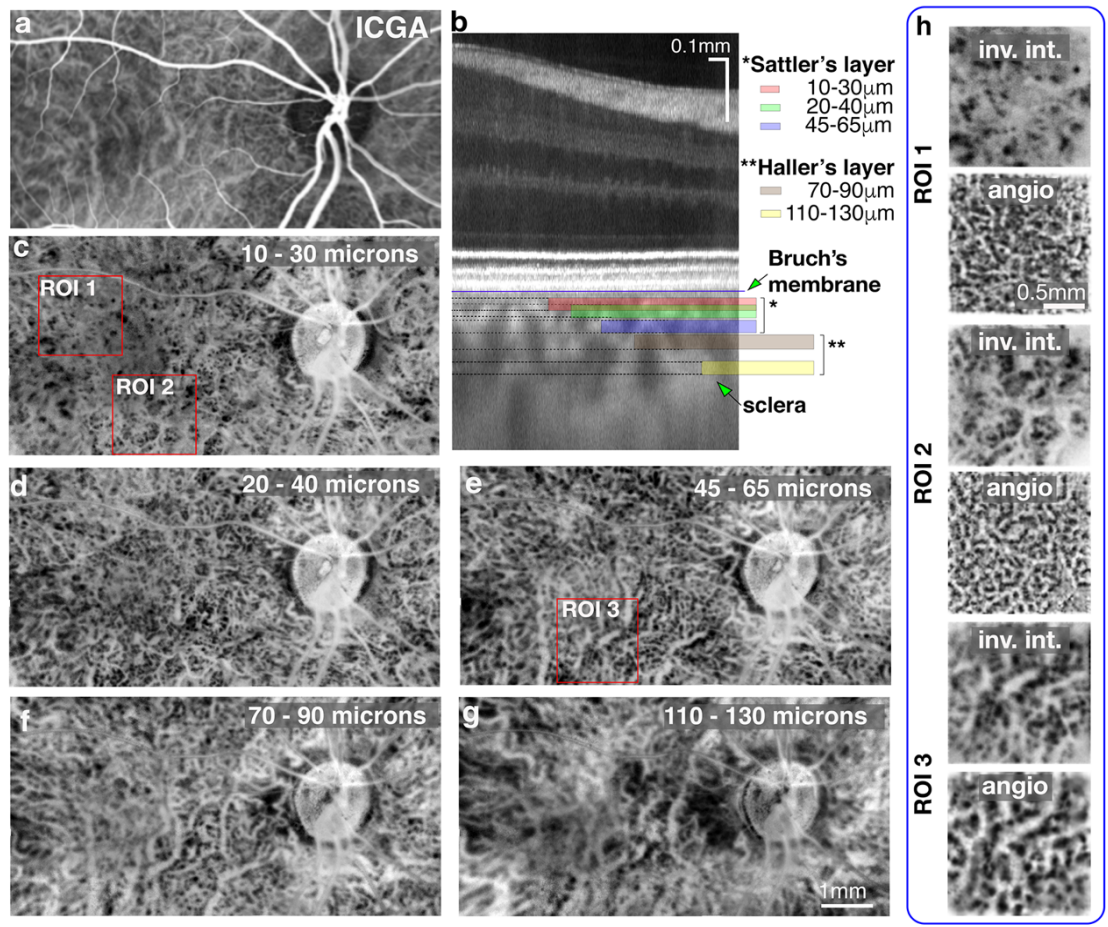

**Figure 3.** STOC-T Imaging of the choroid: **a.** ICG angiography image exposes mostly the large vessels of Haller's layer; **b.** A representative cross-sectional image with the locations of arbitrarily selected projections from the choroid thickness marked by color bars displayed in panels c-g. **c-g.** Contrast-inverted projections derived by averaging STOC-T structural *en-face* images (20 μm - thick) in the indicated axial range below a line demarcating RPE and Bruch's membrane. The images in the range of 10 μm – 70 μm represent the Sattler's layers, and below that the Haller's layer. Vessel diameter can be seen increasing with the imaging depth which correlates with ICG angiography images (Fig. S 13). **h.** Zoomed-in ROI I - ROI III areas marked with red boxes in panels c. and e.; corresponding STOC-T angiography reconstructions are shown in the adjacent panels revealing more details.

The complexity of choroidal tissue makes it optically heterogeneous and difficult to image at different depths. The best currently known method for visualization of choroid vessels is ICG angiography - this method due to its time selectivity allows efficient filtering of signal coming from tissues other than blood vessels and allows observation of choroid projections in short times after contrast administration (see Fig. S 13). However, ICG angiography images show mainly the large vessels of Haller's layer (Figure 3a). The vascular structure of the Sattler's layers is only visible in the area of the optic disc.

By using STOC-T, we can appreciate the richness and complexity of the choroidal vascular system acquired at various depths in the choroid, as shown in Figure 3. In our work, we embraced arbitrary choroidal layer thicknesses and their locations deep within the tissue to show the most representative and non-redundant structures (Figure 3b). STOC-T enables very efficient averaging of volume reconstructions, such that signals with high local dynamics related to blood vessels are clearly hypo-reflective relative to the static part of the choroid that contains connective tissue, neurons and pigments (Fig. S 15). To emphasize the vascular structure, we inverted the contrast of the projection images so that the hypo-reflective vessels became hyper-reflective and the vascular architecture became clearly visible (Figure 3c-g and Fig. S 15).

The images show a large variation in vascular patterns as a function of depths. There is a noticeable difference in the choroidal structure already with a 10 μm change in depth (Figure 3c, d), indicating that the images are not a result of projections happening from the upper layers. The same effect of varying vascular patterns for different depths is observed for angiography images acquired with STOC-T ( Fig. S 14 and Fig. S 15). The STOC-T angiography images display an increased contrast compared to the contrast-inverted projection images, clearly visualizing the vascular system of different choroidal layers, as illustrated in Figure 3h for selected Regions of Interest (ROIs). Therefore, STOC-T angiography images were selected to perform further numerical analyses on various morphometric parameters. It is important to note that both structure (morphology) and angio reconstructions in STOC-T are derived from the same set of data with no modification of the measurement protocol required.

### Choriocapillaris

The choriocapillaris (CC) is a dense vascular monolayer of interconnecting capillaries adjacent to the outer surface of the Bruch's membrane and the RPE cells (Figure 4a). The continuous and broadly anastomosing meshwork of the CC is supported by a rigid framework of connective tissue avascular septa. The thickness of the CC layer ranges from 6.8 μm in the *fovea centralis* to 5 μm 1 mm away from the macula[27]. Together with Bruch's Membrane (BM), which does not have a distinct internal structure in OCT imaging, the CC+BM complex is approximately 12 μm thick and is hyporeflective in OCT cross-sectional images, creating a characteristic demarcation line between the bright layers of the pigment epithelium and the choroidal layers. The blood flow within the CC is supplied by small choroidal vessels which connect to the CC through inlets perpendicular to its plane. The CC bed is responsible for providing oxygen and nutrients to the outer one-third of the retina as well as removing the metabolic waste from the RPE and photoreceptors. The molecular transport between the choroid and retina is facilitated by the fenestrations in the CC endothelium, mostly facing toward the retina. Thus, the CC, RPE and photoreceptor cells form a unitary metabolic complex. This fact contributes to the fundamental role of the CC in the pathophysiology of retinal diseases [28].

Histological studies have shown that CC vessels have a different morphological arrangement in different regions: a dense honeycomb network of freely connected capillaries separated by septa in the macular region, and a polygonal lobular network (Figure 4b) in the equatorial and peripapillary regions. Although the ability to quantitatively image he CC *in vivo* would be of great value, a clinically useful method that reliably provides this information has not been demonstrated to date. Figure 4c shows 3-D rendering of STOC-T data from the choroid area. To better visualize the interrelationship among the 3 different choroid layers, the CC, Sattler's layer, and Haller's layer were highlighted with different colors. This colorized depiction demonstrates that with growing depth the diameter of the vessels increases and their density and degree of branching decreases. Green arrows point at two exemplary locations of Sattler's layer vessels inserting into the plane of the choriocapillaries, approximately perpendicularly to the CC. The white openings correspond to insertions of respective feeding and draining Sattler's vessels into the outer surface of the capillaries (other exemplary locations are marked with red arrows), that indicate the functional lobular architecture. In contrast, the large vessels of Haller's layer heading through the medium vessels towards the openings are marked in yellow.

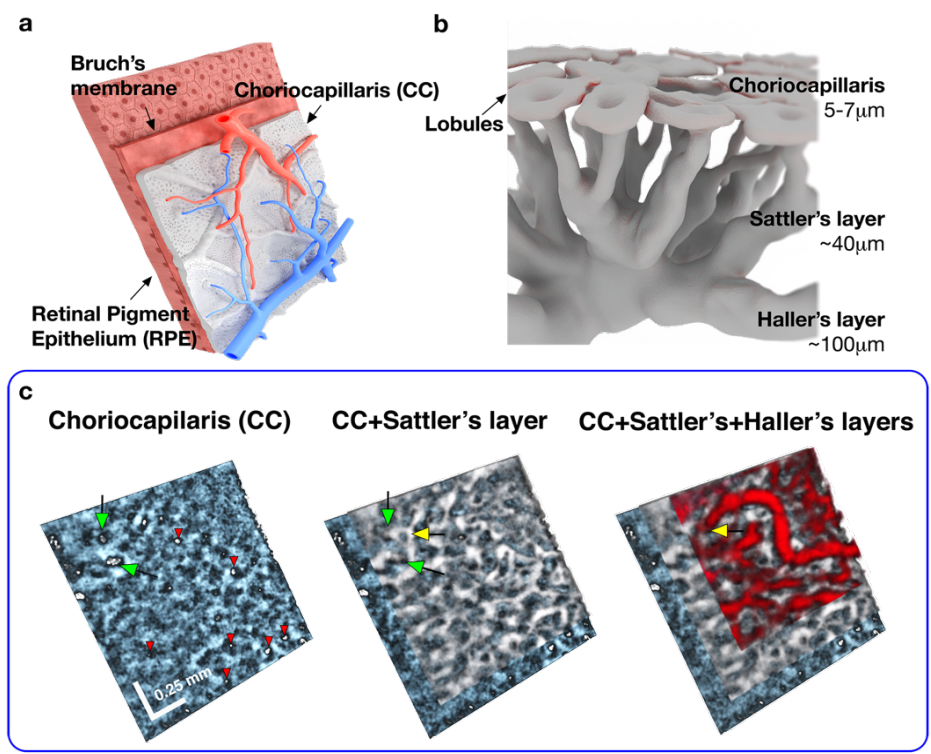

**Figure 4**. STOC-T Imaging of the choroid: **a.** graphics showing the blood circulation system in the choroid supplying the choriocapillaris, relative to the localization of the retinal pigment epithelium as seen from the sclera; **b.** artistic representation of the choroidal layers and choriocapillaris with indicative thicknesses of each layer. The diagram shows the lobular arrangement of the choriocapillaris and the increasing size of the choroidal vessels with depth; **c.** STOC-T imaging of the choroid reveals Inter-space relationships among the retinal vascular layers, rendering three distinct choroidal layers. White bars represent 0.25 mm; Data were rendered in FluoRender free software (http://www.sci.utah.edu/software/fluorender.html).

Figure 5 shows the results of choriocapillaris layer extraction. CC visualization was carried out by using the angio algorithm on 10 µm thick layers. To produce angiography images, the same data sets were used as in previous sections that allowed image stitching to the size of 9mm × 4.6mm. Each 1.7mm × 1.7mm image that constituted the montage was analyzed separately with the angio algorithm. The STOC-T angiography signal coming from the choriocapillaris layer is a relatively week signal that is detected against a high background signal coming from a highly reflecting structure. It, therefore, required additional adaptive histogram equalization [see Extended Data: STOC-T angiography] followed by the blood vessel contrast enhancement using Frangi Hessian filtering (Figure 5d). The ROI areas 1-3 represent individual tiles of the mosaic shown in panel a. To improve visualization of the large number of details in panel a, an image of the mosaic after Frangi Hessian filtering is shown. ROI 2 corresponds to the area located directly in the *Fovea Centralis* zone as indicated with a red dashed circle. To compare the morphology of reconstructed CC layer we show an example of a Scanning Electron Microscope (SEM) (adapted from[29]) on a scale identical to the ROI1-3 (Figure 5). It can be seen that the dense vascular network is reproduced in STOC-T with similar morphological features. In addition, the functional lobular nature of the choriocapillaris structure is manifested by the presence of small vascular subsystems of specific structure (indicated by green arrows), whose recurrence suggests their function of supplying or draining blood from the choriocapillaris (Figure 4a).

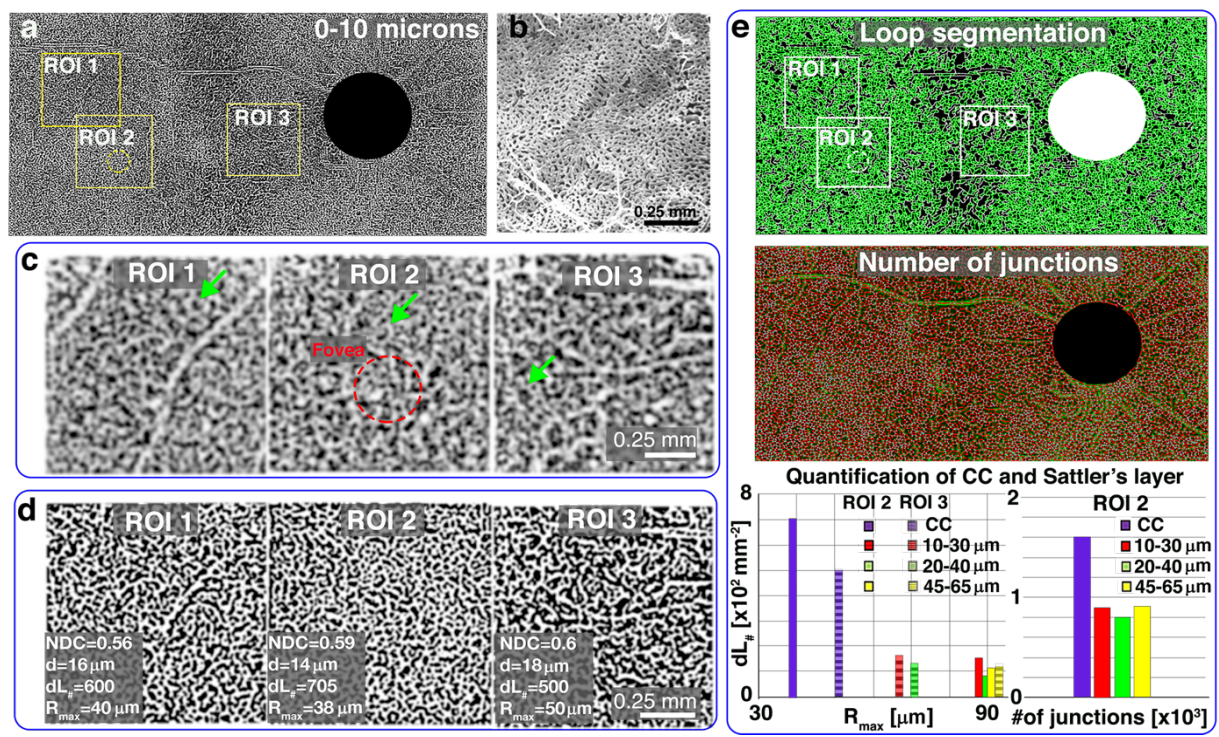

**Figure 5.** Imaging of the capillary lamina of the choroid (choriocapillaris) and its morphometry: **a.** STOC-T angio map (binarized) of the 10 μm - thick layer located next to the RPE, showing the capillary network image after Frangi-Hessian filtering; **b.** SEM image showing the morphology of the choriocapillaris (adapted from [29]); **c.** magnified sections of the STOC-T angio maps (non-binarized version) from the areas highlighted in panel a.; green arrows indicate a small vascular sub-system with a specific morphology recurring in each ROI; **d.** contrast enhanced STOC-T angiography images (binarized) obtained by using the Frangi-Hessian algorithm; **e.** Quantitative analysis of the choriocapillary vascular system by automatic detection of loops and junctions; the graphs show the results of counting loops and junctions for the choriocapillaris and three projections from Sattler's layer, reproduced in Figure 3; NDC - normalized capillary density, d – mean capillary size, $dL_\#$ - local capillary density, $R_{max}$ – maximum radius of loop present in ROI.

Quantification of choriocapillaris morphometry

Unlike existing choroidal imaging modalities, including OCT, *en-face* and angio structure images obtained with STOC-T provide clear contrast and are consistent with each other for any depth of the choroidal layer. This allows quantitative analysis of the acquired data to identify an effective biomarker for assessing disease progression or treatment effects. To this end, we selected methods that quantitatively differentiate the choriocapillary layer according to its location within the posterior pole, and also allow it to be separated from the densely vascularized Sattler's layers. Because of the specific reticular shape of the choriocapillaris and Sattler's layers, we proposed a morphological analysis based on two easily identifiable and robust imaging parameters: the morphology of the loops surrounding the no-flow zones and the number of connections between the vessels (Online Methods).

The number of loops per 1mm$^2$ corresponds to the density of the vascular network and was named here as a local capillary density, $dL_{\#}$. In turn, the size range of the loops indicated the morphological variability of the analyzed vascular area. To characterize this variability, we identified the maximum loop size present in a given area in a significant number of segmented loops; we call this value the maximum loop radius $R_{max}$. The results were collected in a graph (Figure 5e and Fig. S 12), which showed a large disparity between the vessel density in the CC layer relative to the first of the arbitrarily selected depths of the Sattler's layer (a 20 µm slab with a center located 20 µm away from the RPE). This analysis revealed an approximately 5-fold higher capillary density in the choriocapillary layer at a distance of layers only 10 µm apart in depth. The $R_{max}$ value also showed specificity for different ROI locations for the choriocapillaris. According to anatomical knowledge[29], the density of capillary vascular plexuses is the highest and least variable (homogeneous reticular system of vessels) in the area under the fovea centralis; while in the peripheral parts of the retina the system thickens, creating a distinct lobular pattern. In our study, we observed this morphological variability *in vivo* in a non-invasive manner, already documenting a significant variation in $R_{max}$ between the area of the fovea centralis and the area located close to the optic nerve disc.

The mean value of the loop radius in turn allowed us to find the mean value of the capillary size, *d*. This assumption is only valid under the condition that the ratio of the sum of the non-flow areas (the dark regions in the angiography signal) and the sum of the flow areas is constant. The latter value, in turn, is termed elsewhere the normalized capillary density NDC[30], which is approximately 0.6 for all ROIs here. NDC does not give specific readings for different ROI locations in choriocapillaris and therefore does not give a good prognosis for being a sensitive marker for tracking disease progression. However, due to the consistency of the NDC value, we can use the mean radius of the segmented loop to characterize differences in capillary size, *d*. Another way to characterize the morphology of the vascular system is to count all the vascular junctions present in a given area. Figure 5e shows the result of counting junctions in 1mm$^2$ of the choriocapillaris area and in the corresponding areas of different depths of the Sattler's layer. For the choriocapillaris layer about 7300 branches were counted while in the first depth of the Sattler's layer about 4000 branches were found. This value is specific over different depths but is not specific within the choroidal layer.

## Discussion

The results of this study establish that STOC-T is capable of *in vivo* visualization of major retinal and choroidal layers, including that of the choriocapillaris. Conventional - scanning OCT has not yet been able to obtain high detail on the morphological structure of the retina and choroid from a single data set without adjusting the scanning protocol or using sophisticated hardware solutions such as in AO-OCT. Although other wide-field systems, such as time-domain[31] and Fourier-domain[19] full-field OCT can be used for retinal imaging, they have not yet, however, succeeded in imaging the choroid, or even more challenging the choriocapillaris *in vivo*; despite the former technique being crosstalk-free, and the latter being a volumetric (fast) imaging technique. In contrast, STOC-T enables imaging of the choriocapillaris, choroid, and details of ILM, NFL or photoreceptor mosaics by applying various signal and image processing tools. In the attempt to improve retinal imaging with STOC-T, we have previously projected sets of random speckle patterns on the pupil plane of the eye and the retina[23] with the aid of an ultrahigh-speed membrane[22]. However, the achieved axial imaging range was limited since the retinal images contained some depth-dependent modulation artifacts that were difficult to remove computationally. Here, we obtained higher quality images over the larger axial range by using a multimode fiber, with carefully chosen parameters, that was able to project 4 times more speckle patterns within less than 15 nanoseconds (see Crosstalk removal with multimode fiber), which is 3 orders of magnitude faster than it was possible with the membrane[23,24]. The multimode fiber is less expensive, provides more repeatable results and acts as a passive device since it has no moving elements, and thus, is less prone to failure and overall more suitable for use in a clinical environment. The superior speed of the speckle projection provided by the fiber was only partially exploited here and its full potential could be used in the future as it in principle can remove crosstalk within 15 ns. A single retinal volume could be imaged 10 times faster with the current system; *i.e.*, under a millisecond, as limited by the speed of the swept source laser. More decorrelated speckle patterns could be produced by, for example, increasing the core size of the multimode fiber; however, this approach has to be balanced against the depth of field (DOF), which is critical for volumetric imaging with Fourier-domain imaging methods, such as STOC-T. Nevertheless, we find that just ∼400 modes, as produced by 50 μm core multimode fiber, are already sufficient to reduce the crosstalk in retina to negligible levels and maintain substantial DOF of >3 mm, as shown in Fig. S 1(c). Importantly, since such illumination occupies only a fraction of the eye's pupil (∼100 μm), whereas the backscattered light from the retina fills the whole pupil, the current design enables efficient separation of illumination and detection paths. We exploit this arrangement for a lossless and inexpensive implementation of a video-rate display of axial retinal images that can be used as a visual guide for eye alignment (see Online Methods). This alignment technique was valuable here in producing large (montaged) images of the retina and choroid, as it enabled us to quickly find the focus each time a new imaging location had to be imaged. The main 2-D ultrahigh-speed camera cannot be used for generating real-time axial images of the retina due to the limited data transfer rate between the camera and the computer.

*In vivo* imaging of the choriocapillaris layers is very challenging for currently existing techniques, because of the small thickness of choriocapillaris layer, the strong extinction of the optical signal, and the rapid leakage of dye from the fenestration of the capillaries themselves. Consequently, most studies of choriocapillaris function and morphology related to retinal disease etiology and progression have been limited to postmortem histology or indirect observations. The only method that indicates the possibility of *in vivo* imaging of the choriocapillaris is currently OCT. However, the choriocapillaris reconstructions achieved by OCT are too low in contrast to be analyzed quantitatively, as evidenced by the lack of results showing variation in choriocapillary density for different locations in the posterior pole of the eye[32]. The low contrast of OCT reconstructions is due to the presence of speckle noise and the limited lateral resolution of current commercial OCT systems. Also, the depth of the typically analyzed layers is more indicative of one of the first Sattler's layers than of the choriocapillaries themselves[33]. With the size of intercapillary distances (ICDs) ranging from 5-20 μm at the posterior pole, imaging with the 15-20 μm resolution achieved by commercial devices provides rather poor imaging quality. Furthermore, it was recently found that angio-OCT images of the choriocapillaris from different choroidal depths have a similar appearance, suggesting that projection from the choriocapillaris is important in image formation, and may be superimposed on signals from the Sattler's layer, which would explain the observation of a dense network at depths greater than 20 μm from the RPE[34]. In such case, quantitative analysis does not make much sense because there is no specificity between the different layers of the choroid. Slightly better results have been achieved, allowing preliminary morphometry to be performed with AO-OCT[30]; but the high complexity of the system, the small measurement range, and the long measurement time precludes this technique from being widely used in clinical practice. In addition, images from deeper choroid locations have not been shown using this technique, probably for the same reasons mentioned above.

STOC-T provides choroidal images with better image quality compared to conventional OCT, which is somewhat counterintuitive, since conventional OCT uses confocal detection in addition to the temporal gating, which should in principle allow deeper imaging because of better suppression of multiply scattered light. STOC-T acquires higher quality images solely by relying on the temporal coherence gating and phase randomization that remove crosstalk noise. The speckle size is as important as resolution in retinal imaging and so the advantage of STOC-T is not necessarily the achievable imaging depth as such; rather, the key advantage comes from the nearly three-times smaller speckle size (3.3 μm)[23], which does not mask the real retinal/choroidal structures to the same degree as the larger speckles do in conventional OCT. This is because the speckle size in wide-field imaging systems, such as STOC-T, is largely determined by the pupil size and not by the optical aberrations[17], which is in contrast to conventional OCT, where aberrations limit the pupil size (to around 2-3 mm) over which the backscattered light from the retina can be collected through the confocal pinhole (*e. g.*, fiber core). Even though the speckle size can be reduced with Adaptive Optics (AO) implemented with conventional OCT[35], AO-OCT is a cumbersome and complicated technique, with limited field of view and axial range. For example, even though AO-OCT was able to visualize the choriocapillaris and other choroidal layers with 2.4 µm lateral resolution, the field of view was limited to ~0.3 mm × 0.15 mm and the axial range to ~30 µm[30]. In addition, STOC-T can also reduce the speckle contrast[23] by averaging multiple volumes. The images reported here do not show significant speckle contributions because they are under-sampled and appear as single dots (while the resolution is appropriately sampled, speckles are not because the speckle size is 3 times smaller[23]). Another important parameter differentiating STOC-T from conventional OCT is speed and scanning(less) mechanism. Namely, STOC-T is able to acquire 3D volumes of the retina in milliseconds since each pixel of the ultrahigh-speed camera acts as a separate single-point detector enabling simultaneous signal acquisition from parallel retinal locations. This parallelization makes mechanical scanning obsolete and, subsequently, volumetric signal detection faster compared to the conventional OCT that employs a single-point detector and galvanometer scanners. The volumes are normally averaged to increase the image contrast (SNR) but the effective A-scan is still very high – 3 MHz when 10 volumes are averaged. Such images are also devoid of point-scanning artifacts encountered in conventional OCT. STOC-T can computationally recover aberrated signal and restore, for example, details of the photoreceptor layer, which is possible despite phase randomization[24] with the multimode fiber. Not only does STOC-T restore resolution, but it also utilizes an available photon budget more efficiently, since aberrated signal is normally rejected by the pinhole in the conventional OCT. This is important when the amount of light that can be safely launched into the eye is restricted or when using limited-power light sources.

To summarize, we have advanced STOC-T technology with a series of developments that improve quality of the images acquired with this technique and make it more suitable to use in a clinical environment. The advancements help us to generate high quality retinal and choroidal images with a simple and robust optical layout. Specifically, we employ a more efficient method to reduce crosstalk in retinal images that permits visualization of choroid and choriocapillaris with a better contrast and imaging depth. We have also implemented a real-time preview mode that facilitates quick alignment of a patience's eye for the imaging procedure, which is essential for clinical deployment and imaging over a large field of view.

## Online Methods

### STOC-T system

STOC-T is a three-dimensional imaging technique that can achieve micrometer resolution deep in tissue. Its principles rely on low coherence interferometry that acquires multispectral interferometric images with a camera and a swept (tunable) light source (Extended Data Fig. S 1). It subsequently produces 3D data (volumes) of retina through Fourier transformation. The STOC-T systems for retinal imaging, shown in Fig. S 1(a), broadly consisted of a tunable laser source (Broadsweeper BS-840-2-HP, Superlum), an interferometer and an ultrahigh-speed 2-D camera (Fastcam SA-Z, 2100K-M-16GB, Photron). It is difficult to precisely determine lateral resolution in the eye, so we estimated it by imaging an USAF resolution target in the reference arm, which gave an estimate of $\sim 7 \mu m$. Measured axial resolution was 5 μm. Spatial coherence gating was measured to be at least 2.9 mm (see in Fig. S 1). The achieved field of view was 1.7 mm x 1.7 mm for 512 x 512 images. It was acquiring retinal volumes with ~ 100 Hz rate.

### Crosstalk removal with multimode fiber

We employed a 300 meter-long step-index multimode fiber (Thorlabs, FG050LGA) with silica core of 50 μm and *NA* of 0.22 to deliver the laser radiation to the optical system. The fiber was connectorized with FC/APC connectors to prevent back reflections from the tip of the fiber back to the laser cavity. The fiber breaks down a spatially coherent laser illumination at 840 nm into N~800 spatial modes (eq. 1 in Extended Data), with each of them acting as an uncorrelated speckle pattern. To make sure that speckle patterns do not interfere with each other but add incoherently, a 100 meter-long piece of the fiber must be used (eq. 4 in Extended Data). In practice, we choose longer fiber (of 300 meters) because we find that the fiber-induced speckle contrast is lower which could be explained by the presence of nearly degenerate modes that travel closely and might be responsible for the residual coherence between some modes[36,37]. The time separation, δT in such length between the fastest and the slowest mode can be calculated (eq. 2 in Extended Data) to be δT = 15 ns, meaning that all 800 speckle patterns should come out within this time. In practice, the number of modes that we are able to excite in the fiber was estimated to be ~ 400 from measured contrast (of 5%) and using eq. 5 (in Extended Data). Since crosstalk reduction scales as ~ $N^{1/2}$ [38,39], it should therefore reduce crosstalk 20-fold. So, to remove crosstalk in the images to insignificant level it is not necessary to make imaging fully incoherent, by filling the whole pupil plane; filling a tiny fraction can achieve satisfactory results.

### Video-rate retinal axial preview

To implement a video-rate display, we designed an illumination path that incorporated a small pick-off mirror and a line (1-D) camera (Alkeria, Necta N4K-7-F), as shown in Fig. S 1(a), that is synchronized with the 2-D camera (to run at 60 kHz acquisition rate). The pick-off (rod) mirror was placed in the conjugate pupil plane in the illumination path that efficiently separated the illumination and detection paths because, on one hand, it was big enough to reflect all the illumination light towards the eye but, on the other hand, was small enough to let essentially all the backscattered light from the retina to the line camera. Therefore, the pick-off mirror effectively opened a second detection channel without compromising the first (main) being used by the 2-D camera. This was based on the fact that the illumination provided by the core of the multimode fiber was occupying only ~100 μm of the pupil, whereas the backscattered light filled it completely. Since in practice we use a pick-off mirror of $1mm$ and we place it in a more-accessible twice bigger pupil-conjugate plane (Fig. S 1a), it results in insignificant $(1mm/12mm)^2 = 0.7\%$ attenuation of the backscattered light (assuming homogenous distribution of the backscattered light across the pupil). The reference mirror was tilted at least by 0.8°, as shown in Fig. S 1a, for the reference bean to pass around the pick-off mirror. A single B-scan is generated by first capturing a single interferometric line (say, along $x$ dimension) with the line camera for each wavelength, λ. When the lines are combined into an xλ image, a Fourier transform is performed along λ for each $x_n$ position, where n is the pixel number on the camera. Effectively, we are only detecting a narrow strip of the 2D signal that is falling onto the line camera. The line camera acquired a single axial image of the retina every few milliseconds that helped with eye alignment and selection of imaging location.

Data Analysis

Acquired stack of multispectral interferometric images (x, y, λ) from the retina with the 2-D camera are converted to 3D rendering (volumes) of the retina in postprocessing by the following procedure (Extended Data Fig. S 2). First, we subtract the mean from each line (along λ, A-scan). The resulting signal is then apodized with Hamming window, and converted from spectral to depth domain (λ to z) by the Fast Fourier transformation (FFT). The achieved volumes (x, y, z) are then spatially filtered along depth to remove low-frequency speckles arising from the multimode fiber[40]. We then correct the data for chromatic dispersion mismatch between interferometer arms and the eye's possible axial motion during the laser sweep. We do so by using an iterative approach, which corrects the phase of each B-scan (xz or yz) until the image sharpness is optimized (see in Extended Data: Digital Aberration Correction). To quantify B-scan sharpness we used kurtosis. Subsequently, we corrected the volumes for depth-dependent defocus by using sub-apertures approach. Higher-order aberrations were then corrected iteratively by optimizing the *en face* (xy) images at selected depths (see in: Extended Data: Digital aberration correction). The aberrations were parameterized using 21[st] Zernike polynomials, excluding piston, tilt, and tip. The corrected volumes were then registered using correlation-based approach, and aligned volumes were upsampled with bicubic interpolation, integrated incoherently, and flattened with respect to the photoreceptor layer, to produce STOC-T volumes with the structural representations of the retina. The registered volumes, before averaging, were also employed to determine angiography volumes. The consecutive pairs of complex-valued volumes were used to calculate complex differential variance (CDV). The resulting CDV volumes were then convolved with a three-dimensional kernel (Hamming 7x7x21), and averaged. The *en face* angiographic projections were achieved from those averaged volumes by selecting the pixels with intensities above a pre-defined threshold. Finally, the *en face* angiographic images were enhanced through adaptive histogram equalization.

### Quantification of choriocapillaris morphometry

To quantify choriocapillaris, adaptive histogram equalization was performed to optimize image contrast. The *en face* image was processed to remove noise, and the Hesian-based Frangi method with sigma = 1 was used to enhance the blood vessels. The image was then binarized using the Phansalkar method with a local window radius of 15 pixels[41]. Axially flattened binarized maps were processed to identify local vascular cross-sections and calculate local maxima, and were then masked for skeletonizing[42]. Loop analysis allowed us to indicate the range of capillary sizes and density for different locations, both laterally and in depth, for the CC and Sattler's layers at different depths. The analysis was performed for STOC-T angiography reconstructions within areas of 1mm$^2$ (ROI1 ROI2 ROI3 shown in Figure 5a) for the CC, and corresponding reconstructions for Sattler's layers with identical filtering and segmentation parameters (see Extended Data: Fig. S 12). The total number of loops in a given area was counted and the distribution of the number of loops as a function of their size was analyzed.

Quantitative evaluation of angiography including spatial parameters related to branching index, such as total number of junctions and density of junctions in vessels, is performed using a semi-automated computational tool[43]. The evaluation parameters were chosen for each image to ensure optimal vessel detection. These metrics are used to characterize the heterogeneity of the vessel and highlight the variability in each layer: Haller's layer, Sattler's layer, and the choriocapillaris. (The vessel outline is shown in yellow, the vessel skeleton in red, and the junctions are represented by blue dots).

### In vivo measurements

The study was conducted at the International Center for Translational Eye Research at the Institute of Physical Chemistry, Polish Academy of Sciences. The study protocol was approved by the ethics committee at the Collegium Medicum of Nicolaus Copernicus University (KB 311/2018). Written informed consent was obtained from a healthy 45-year-old volunteer who participated in the study. A translation stage-mounted chinrest was used to align the head of an individual. The measurement procedure requires the acquisition of multiple volumes located at different positions in the retina. We acquired 32 volumes at each position, which were used in signal processing to produce a final high-quality data set. To obtain proper control over the orientation of the eye at a given position, we used a "ship game" board that was displayed in front of the volunteer's eye. Once the desired eye starting position was found, using high-speed single-axis viewing on a line camera, we measured different locations by instructing the volunteer which part of the board to look at with the eye while keeping the head position constant. Real-time axial images of the retina were used as a visual guidance to (I) find signal from the retina by axially translating the chin rest until optical paths between the sample and reference arms were matched and (II) to optimize the signal in terms of axial image sharpness by adjusting the eye lens (L4 in Fig. S 1(a)) which allowed us to account for the refraction error in the eye. The acquisition interval between two consecutive positions is about 20 sec, which gave a total measurement time of about 10 min.

Control Software

The software to control the experiment was developed in LabView. Three modes of operation are available. The first one, preview mode, enables acquisition of B-scans with the line scan camera. We use this configuration for initial eye alignment. The second mode, called B-scan mode, allows visualization of b-scans acquired with the ultrafast camera. However, this b-scan is in a direction perpendicular to the one acquired with the line camera. In this mode, the line camera is not used. With this additional information, the person taking the measurement can precisely position the eye in the desired starting position. The third mode, measurement, is the most important one, and uses the detection potential of the ultrafast camera. Switching it on initiates ultrafast acquisition of a predefined number of volumetric data sets synchronized in time with laser sweeping. After completing the data transfer, the software returns to the last used mode (b-scan or preview mode). The acquisition is preceded by the configuration of the ultrafast camera: setting the predefined ROI on the sensor, frame rate, exposure time and camera trigger mode. In this configuration phase, the software defines the acquisition control waveforms and provides fixed acquisition intervals for the corresponding volumes. The software provides communication with the sweeping laser, ultrafast camera, line camera, and National Instruments I/O card responsible for the triggering of data acquisition to implement those functionalities. The trigger initiates a sweep of the laser and times data acquisition with a line camera in the preview mode. In b-scan and measurement modes, the trigger controls the sweep and ultrafast camera.

Acknowledgements
Foundation for Polish Science (MAB/2019/12); National Science Center (NCN, 2016/22/A/ST2/00313); European Union's Horizon 2020: research and innovation programme (666295); Polish Ministry of Science and Higher Education (2016-2019 int. co-financed project); European Regional Development Fund (project No. 01.2.2-LMT-K-718-03-0093) under grant agreement with the Research Council of Lithuania (LMTLT).
The International Centre for Translational Eye Research (MAB/2019/12) project is carried out within the International Research Agendas Programme of the Foundation for Polish Science, co-financed by the European Union under the European Regional Development Fund.